# Acoustic Holographic Rendering with Two-dimensional Metamaterial-based Passive Phased Array

Yangbo Xie[1]*, Chen Shen[2]*, Wenqi Wang[1], Junfei Li[1], Dingjie Suo[2], Bogdan-Ioan Popa[1], Yun Jing[2]††, and Steven A. Cummer[1]†

[1]Department of Electrical and Computer Engineering, Duke University, Durham, North Carolina 27708, USA

[2]Department of Mechanical and Aerospace Engineering, North Carolina State University, Raleigh, North Carolina 27695, USA

* These authors contributed equally to this work

†Corresponding authors:
cummer@ee.duke.edu;
yjing2@ncsu.edu;

## Abstract

Acoustic holographic rendering in complete analogy with optical holography are useful for various applications, ranging from multi-focal lensing, multiplexed sensing and synthesizing three-dimensional complex sound fields. Conventional approaches rely on a large number of active transducers and phase shifting circuits. In this paper we show that by using passive metamaterials as subwavelength pixels, holographic rendering can be achieved without cumbersome circuitry and with only a single transducer, thus significantly reducing system complexity. Such metamaterial-based holograms can serve as versatile platforms for various advanced acoustic wave manipulation and signal modulation, leading to new possibilities in acoustic sensing, energy deposition and medical diagnostic imaging.

Holography is a technique to record and reconstruct the complete information of wave fields. While optical holograms have been widely applied in virtual reality displays, data storage, sensing, and security printing[1]. Acoustic holograms, on the other hand, are relatively less developed compared to their electromagnetic counterparts in terms of applications. One major restricting factor is the limited acoustic properties that natural or traditional materials can offer. To date, most acoustic holographic reconstruction techniques[2] rely on phased arrays with large numbers of active elements, requiring sophisticated phase shifting circuits, large power consumption and careful calibration and tuning.

Metamaterials, a family of artificial materials with engineered micro-structures, can provide flexible and unusual effective material properties thus offer new possibilities of designing and fabricating holograms. Recently a variety of electromagnetic holograms based on metamaterials have been experimentally demonstrated[3-6]. While many acoustic metamaterial-based unconventional wave controlling and sensing devices have been reported[7-16], acoustic holograms based on metamaterials, however, have not been reported to date.

In this work, we demonstrate the designs and experimental realizations of two acoustic metamaterial-based holograms: one projects a letter 'A' pattern on the image plane, while the other focus energy onto multiple circular spots of different sizes. The design was aided by an iterative hologram generation and optimization algorithm, and also verified with two numerical simulation tools. The desired phase patterns were physically realized with labyrinthine acoustic metamaterial unit cells tailored for the holograms. The experimental testing of the holographic reconstructions was carried out in an anechoic chamber. Good agreement was found between the measured holographic reconstruction and the designed patterns. Such acoustic metamaterial-based holograms are not only direct acoustic transposition of optical computer generated holograms, but pave the way to advanced acoustic wave manipulation and complex field reconstruction using passive acoustic metamaterials without the need of phase-shifting circuitry and transducer arrays.

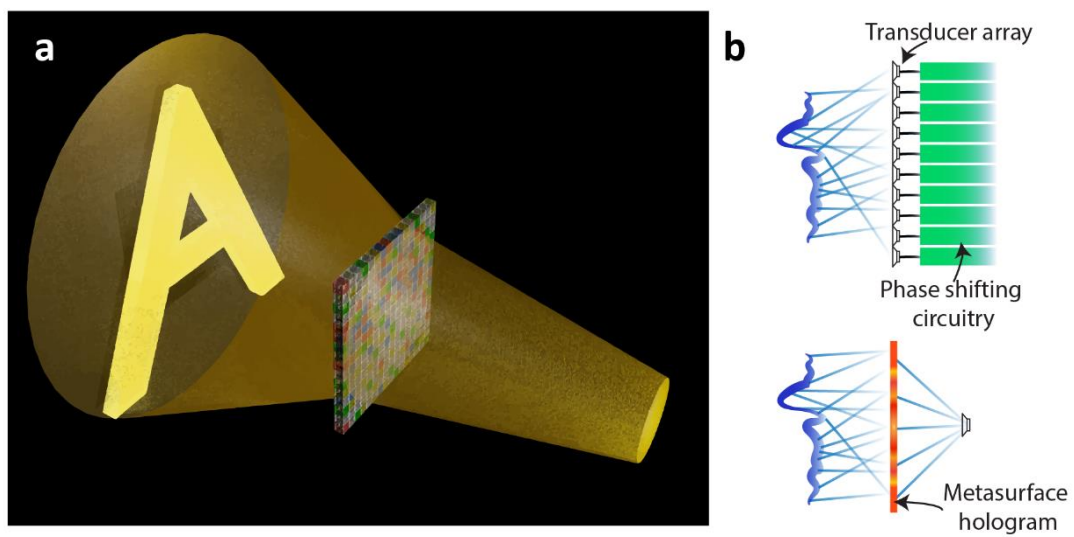

**Figure 1 | Schematic of holographic rendering with an acoustic metamaterial-based hologram. a**, A schematic showing the holographic rendering of a letter 'A' with the passive metamaterial-based hologram. **b**, Comparison between holographic rendering with an active phased array and a hologram

made of metamaterial-based passive phased array, which eliminates cumbersome phase-shifting electronics and a large number of transducers.

**Results**

A hologram can be realized with pixel-by-pixel modulation of the phase or/and the amplitude on the acoustic waves. Here we focus on phase holograms whose transmission amplitude is assumed to be uniform over the plane of the hologram. The Weighted Gerchberg-Saxton (GSW) algorithm[17] is modified and used here to generate the optimal phase distribution of the hologram. A spatial impulse response-based acoustic field simulation tool, Field II, is employed to calculate the projected field on the image plane and integrated with the GSW algorithm to perform hologram optimization. By optimizing the spatial pattern of the amplitude on the image plane iteratively (details in Supplementary Methods), the optimal phase distribution needed for certain reconstruction is obtained. Finally, by compensating for the phase differences from the point source to the pixels of the hologram, the phase delay for all the pixels can be calculated.

The optimized design was verified with two numerical simulation tools: one is the Angular Spectrum approach (ASA), and the other is a Finite Element Method (FEM)-based full wave simulation package COMSOL Multiphyiscs. ASA is a well-established technique and is able to accurately predict the sound field at a distance away from the initial plane. The essential idea of ASA involves decomposing the initial pressure field into plane waves with different wave-vectors and subsequently propagating these plane waves using analytical formula (details in Supplementary Methods). The simplicity of ASA and the fact that it uses fast Fourier transform render it highly efficient for predicting the holographic reconstruction at different depths and excellent agreement was found between the predictions by ASA and the FEM method (e.g., see Fig. 2e). Therefore, we employ ASA as our primary simulation tool for comparison with the measured results.

To realize the phase delay profile required by the hologram, we designed a set of 12 optimized labyrinthine unit cells with gradient transmission phase delays and relatively constant transmission amplitudes (see Supplementary Figure S1 for a more detailed characterization of this set of unit cells). Labyrinthine unit cells are a family of geometry-based non-resonant metamaterials that have been proposed recently[18-20]. They have been shown to possess various attractive features for designing transmissive or reflective acoustic metasurfaces[11, 21]: namely precise phase control, high transmission or reflection, and relatively broad bandwidth due to their non-resonant nature. The set of 12 unit cells, as shown in Fig. 2c, is designed for 4000 Hz covering 180 degrees of relative phase delay. Two layers of unit cells achieve a complete angular coverage of 360 degrees of relative phase change across the hologram. The unit cells are fabricated with acrylonitrile butadiene styrene (ABS) plastics using the fused filament fabrication (FFF) 3D printing technology. The photo of a fabricated hologram is shown in Fig. 2b. Note that the color of the unit cells are randomly selected during the 3D printing process and does not reflect the types of the unit cells.

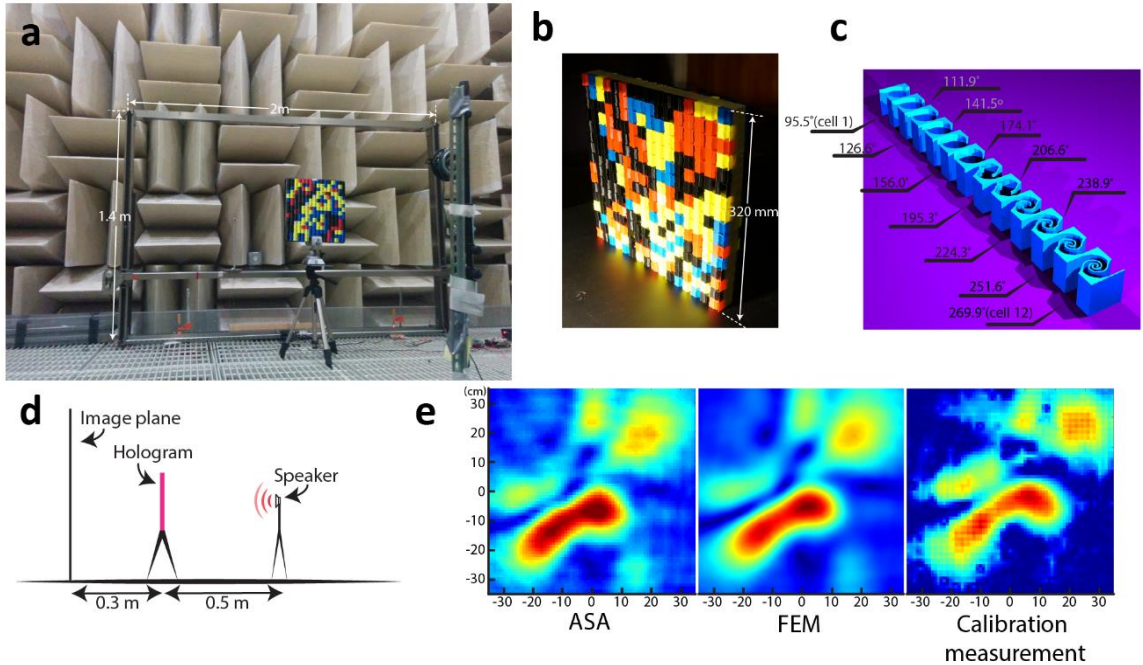

**Figure 2 | Experimental setup for measuring the holographic rendering of the acoustic metamaterial-based holograms. a**, A photo showing the experimental setup. A scanning stage inside an anechoic chamber is employed for mapping the fields at certain depths. **b**, The photograph of a fabricated hologram that is composed of 512 unit cells. **c**, The set of 12 constituting unit cells used to achieve gradient phase delays. **d**, The layout of the holographic reconstruction system. **e**, The simulated results of a random calibration hologram with ASA and Finite Element Method (FEM)-based full wave simulation, as well as the corresponding measurement.

The experiment setup to verify the proposed metamaterial hologram is depicted in Fig. 2a and 2d. The hologram is mounted securely in an aperture in a large sheet of hard paperboard. Since the acoustic impedance of the paperboard is much larger than that of air, it is assumed to be acoustically rigid and prevents sound from bypassing the hologram. To verify our hologram design scheme and the experimental setup, we measured first a randomly patterned hologram and compare the measurements with two aforementioned simulation tools. Excellent agreement was found between the measured field at 30 cm and those with COMSOL simulation as well as Angular Spectrum calculation, as shown in Fig. 2e. The agreement demonstrates that the design method is effective and the verifying simulation tools can faithfully predict the projected field patterns to the extent of even some smaller features.

To demonstrate the capability of our proposed metamaterial-based hologram for creating acoustic 'illusions' for imagers, we designed a hologram that projects the amplitude pattern of the letter 'A' in an image plane 300 mm behind the hologram. Fig. 3a shows the desired pressure amplitude pattern on the image plane. Fig. 3b compares the ideal phase, which is calculated with the GSW algorithm, and the measured phase right behind the hologram. The measured phase agrees well with the desired one, with some small discrepancies that are likely caused by minor fabrication defects. Fig. 3c presents the simulated amplitude pattern on the image plane according to ASA and the measured pattern, where an 'A' is legible. The reconstruction quality can be further improved by using larger holograms with more pixels. We also note the reconstruction of the design pattern preserves over a relatively broad bandwidth. From 3750 Hz to 4500 Hz, the measured amplitude patterns on the image plane possess good consistency with the designed 'A' pattern (see Supplementary Figure S4 for the measured field patterns at these frequencies), indicating an operating bandwidth of more than 18.75% of the central frequency of 4000 Hz. The relative broad bandwidth is caused by the non-resonant nature of the constituting labyrinthine unit cells.

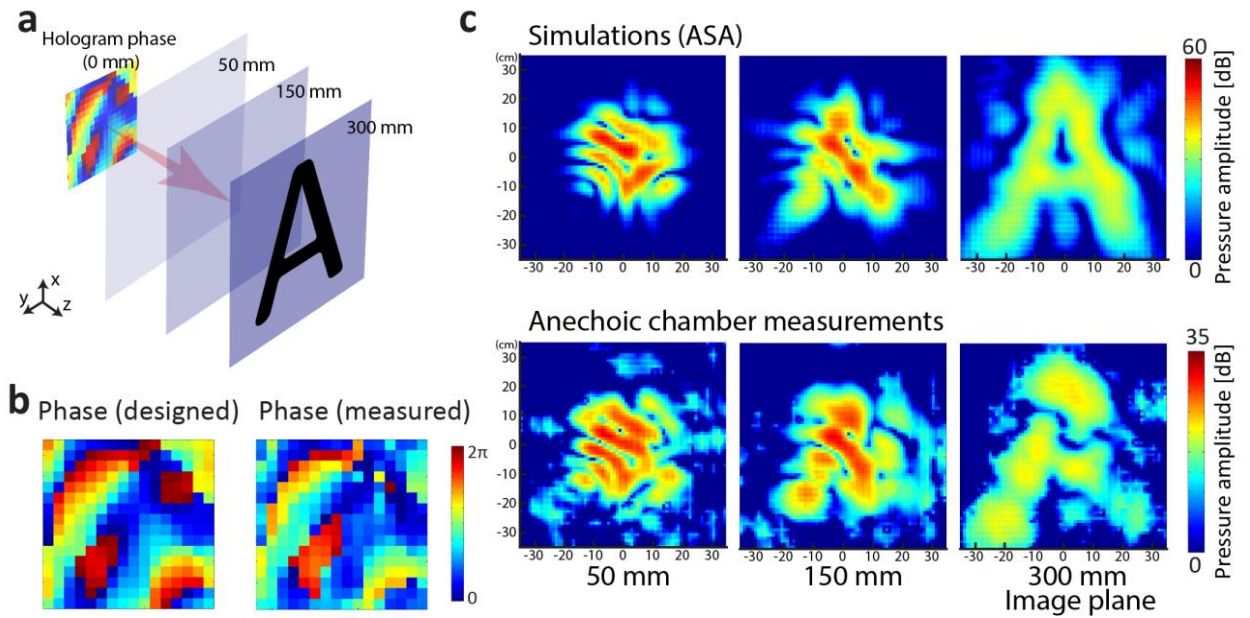

**Figure 3 | Experimental results of the holographic reconstruction of a letter 'A'. a**, The desired pattern to be reconstructed at the depth of 300 mm. **b**, The ideal phase pattern designed with GSW method and the actually measured phase pattern immediately behind the hologram. **c**, The simulated field patterns (amplitude) at three representative depths compared to those actually measured.

Besides projecting a complex pattern, here we also demonstrate that the proposed metamaterial-based holograms can be used as advanced holographic lenses. Acoustic lensing techniques are indispensable for imaging and energy deposition with acoustic waves. Conventional acoustic lens designs generally rely on Lens-maker's formulas or diffraction controls[22] and are thus limited for designing advanced lenses with multiple tailored wavefront-shaping characteristics. The design of holographic lenses, in contrast, brings about unprecedented degrees-of-freedom in designs to create lensing with complex focal patterns. The focal spots are treated as the projected pattern and similar procedures can be followed by that for designing a general computer generated hologram. Here we demonstrate that the metamaterial-based holograms can be used to achieve such a holographic lensing. Starting from a desired power distribution pattern of multiple focal points on the imaging plane (e.g., three focal spots with different sizes as shown on the focal plane in Fig. 4a), we can obtain an optimized phase distribution for the lens using the same GSW algorithm. Fig. 4a shows the isosurface of power and the projected power pattern of 3 energy hot spots on the focal plane (300 mm behind the hologram). Fig. 4b compares the simulated (with ASA) and the measured power distribution, where three focal points are clearly resolved at the expected locations with the one on the bottom right having the smallest size. We also observed relatively broad bandwidth for such lensing effects. The measured focal patterns remain relative constant over the bandwidth from 3750 Hz to 4500Hz, and all three desired focal spots are clearly identifiable at the designed locations (see Supplementary Figure S5 for the measured focal patterns at these frequencies). Such advanced tailored control over multiple focal spots with passive metasurface holograms open the door to a large variety of potential applications, such as wireless power transfer[23], medical therapy[24] and neural engineering[25].

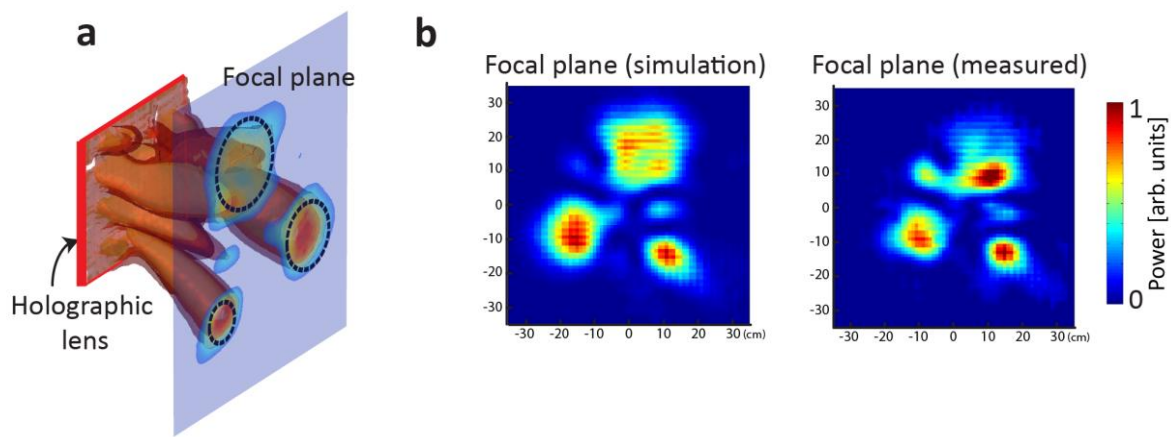

**Figure 4 | Experimental results of the holographic lensing of three circular spots of different diameters. a**, Illustration of the multiple focus lensing with the holographic lens. The isosurface of power and the projected power pattern on the focal plane (300 mm) is calculated with FEM-based full wave simulation. The three dashed circles with different radii are the designed focal spots. **b**, The simulated and measured power distribution patterns at the depth of 300 mm.

**Discussion**

While active phased array-based holographic rendering methods allow real-time adaptive control over each transducer element, the passive holograms we demonstrated here are advantageous for applications where simplicity, robustness and small volume are preferred over adaptive control. For example, our hologram, which acts as a plane to adjust the phase of the impinging sound, can be adapted to compensate for the phase aberration of human skulls for transcranial beam focusing. This could enable precise focusing for brain therapy or stimulation without over a thousand transducer elements and the same number of dedicated driving channels[26], both of which are very expensive and challenging to construct. Since the desired phase correction and the targeted area in the brain can be per-determined, there is no need to adjust the sound field/hologram in real-time. For different patients, the hologram can be reassembled for achieving the specific phase correction and focal patterns. Another example where non-adaptive holographic rendering of acoustic field can be useful is the acoustic tweezers for trapping and manipulating microparticles[27, 28].

Although in this paper we focus on phase-only modulations, complete control over both amplitude and phase is possible using labyrinthine metamaterials[20, 21]. With advanced manufacturing technologies, the proposed holograms can also be scaled down to the ultrasonic range. The underwater version of our proposed metasurface hologram would involve more complications but in principle should be possible. We expect that the proposed metamaterial-based holograms can become a platform for a large variety of wave-based signal processing and modulation functionalities.

To conclude, we have designed, fabricated, and tested two acoustic metamaterial-based holograms. We expect that the results of this paper will open a new realm of holographic acoustic wave manipulation with metamaterials or metasurfaces. The passive holograms will not only be useful for sound field reconstruction, but may also be valuable for various advanced functionalities, such as creating acoustic mirages[29, 30] or illusive objects[31, 32], generating radiation force to levitate and manipulate objects[28], serving as wave-based analog computational platforms where acoustic signals can be processed directly without digitization[33]. Such holograms can also in principle be applied to the ultrasonic regime, to achieve multi-focal lensing for targeted drug delivery and noncontact tactile display and precise neural stimulation[25].

## Method

### Three-dimensional field mapping in an anechoic chamber

The acoustic source is a loudspeaker sending out Gaussian-enveloped sinusoidal pulses with a center frequency at 4000 Hz towards the hologram. The field mapping measurements were performed in an anechoic chamber. A Micro-Electro-Mechanical Systems (MEMS) microphone (ADMP401) was mounted on a two-dimensional linear scanning stage to sweep through the spatial grid on the plane of a certain depth, then the relative distance along z direction between the hologram and the measurement plane was carefully adjusted manually to obtain the mapped field on the plane at another depth. The acoustic pressure at each point is averaged from five independent measurements to reduce the noise.

### Numerical Simulations

Three numerical tools were employed in the processes of designing and verifying the holograms. In the initial designing stage, a spatial impulse responses-based package, Field II, was used to implement the GSW hologram generation and optimization algorithm. Besides, Finite Element Method-based full wave simulation (with a commercial package COMSOL MULTIPHYSICS 5.1 and its Pressure Acoustics module) was used to verify the computer generated holograms. The results of the FEM-based simulation were also compared with Angular Spectrum approach-based calculation using a randomly patterned calibration hologram. Excellent agreement was found between the simulation results obtained from the FEM-based simulation and that from the ASA.

## References

1. Collier, R. *Optical Holography* (Academic Press, New York, 2013).
2. Rossing, T. D. (Ed.). *Springer Handbook of Acoustics* (Springer, New York, 2015).
3. Larouche, S., Tsai, Y. J., Tyler, T., Jokerst, N. M., & Smith, D. R. Infrared metamaterial phase holograms. *Nat. Mater.*, **11**, 450-454 (2012).
4. Ni, X., Kildishev, A. V., & Shalaev, V. M.. Metasurface holograms for visible light. *Nature Commun.* **4**, 2807 (2013).
5. Huang, L. et al. Three-dimensional optical holography using a plasmonic metasurface. *Nat. Commun.* **4**, 2808 (2013).

6. Huang, K. et al. Silicon multi-meta-holograms for the broadband visible light. *Laser Photon. Rev.* **10**, 500-509 (2016).
7. Li, J., Fok, L., Yin, X., Bartal, G., & Zhang, X. Experimental demonstration of an acoustic magnifying hyperlens. *Nat. Mater.* **8**, 931–934 (2009).
8. Zhang, S., Xia, C., & Fang, N. Broadband acoustic cloak for ultrasound waves. *Phys. Rev. Lett.* **106**, 024301 (2011).
9. Lemoult, F., Kaina, N., Fink, M., & Lerosey, G. Wave propagation control at the deep subwavelength scale in metamaterials. *Nat. Phys.* **9**, 55-60 (2013).
10. Wang, P., Casadei, F., Shan, S., Weaver, J. C., & Bertoldi, K. Harnessing buckling to design tunable locally resonant acoustic metamaterials. *Phys. Rev. Lett.* **113**, 014301 (2014).
11. Cheng, Y. *et al*. Ultra-sparse metasurface for high reflection of low-frequency sound based on artificial Mie resonances. *Nat. Mater.* **14**, 1013-1019 (2015).
12. Fleury, R., Sounas, D., & Alù, A. An invisible acoustic sensor based on parity-time symmetry. *Nat. Commun.* **6**, 5905 (2015).
13. Brunet, T. *et al.* Soft 3D acoustic metamaterial with negative index. *Nat. Mater.* **14**, 384-388 (2015).
14. Xie, Y. *et al.* Single-sensor multispeaker listening with acoustic metamaterials. *Proc. Natl. Acad. Sci.* **112**, 10595-10598 (2015).
15. Li Y., Jiang X., Liang B., Cheng J. C. & Zhang L. K. Metascreen-based acoustic passive phased array. *Phys. Rev. Applied* **4**, 024003 (2015).
16. Ma, G., & Sheng, P. Acoustic metamaterials: From local resonances to broad horizons. *Sci. Adv.* **2**, 2 (2016).
17. Di Leonardo, R., Ianni, F., & Ruocco, G. Computer generation of optimal holograms for optical trap arrays. *Opt. Express* **15**, 1913-1922 (2007).
18. Liang, Z. & Li, J. Extreme Acoustic Metamaterial by Coiling Up Space. *Phys. Rev. Lett.* **108**, 114301 (2012).
19. Xie, Y., Popa, B.-I., Zigoneanu, L. & Cummer, S. A. Measurement of a broadband negative index with space-coiling acoustic metamaterials. *Phys. Rev. Lett.* **110**, 175501 (2013).
20. Xie, Y., Konneker, A., Popa, B-I. & Cummer, S. A. Tapered labyrinthine acoustic metamaterials for broadband impedance matching. *Appl. Phys. Lett.* **103**, 201906 (2013).
21. Xie, Y. *et al*. Wavefront modulation and subwavelength diffractive acoustics with an acoustic metasurface. *Nat. Commun.* **5**, 5553 (2014).
22. Azhari, H. *Basics of biomedical ultrasound for engineers* (John Wiley & Sons, Hoboken, 2010).
23. Roes, M. G., Duarte, J. L., Hendrix, M. A., & Lomonova, E. A. Acoustic energy transfer: a review. *IEEE Trans. Ind. Electron.* **60**, 242-248 (2013).
24. Hill, C. R., Bamber, J. C., & Haar, G. (Eds.). *Physical Principles of Medical Ultrasonics* (John Wiley & Sons, Chichester, 2004).
25. Hertzberg, Y., Naor, O., Volovick, A., & Shoham, S. Towards multifocal ultrasonic neural stimulation: pattern generation algorithms. *J. Neural. Eng.*, **7**, 056002 (2010).
26. Pajek, D., & Hynynen, K. The design of a focused ultrasound transducer array for the treatment of stroke: a simulation study. *Phys. Med. Biol.* **57**, 4951 (2012).
27. Hong, Z., Zhang, J., & Drinkwater, B. W. Observation of Orbital Angular Momentum Transfer from Bessel-Shaped Acoustic Vortices to Diphasic Liquid-Microparticle Mixtures. *Phys. Rev. Lett.* **114**, 214301 (2015).

28. Marzo, A. *et al.* Holographic acoustic elements for manipulation of levitated objects. *Nat. Commun.* **6**, 8661 (2015).
29. Layman, C. N., Naify, C. J., Martin, T. P., Calvo, D. C., & Orris, G. J. Highly anisotropic elements for acoustic pentamode applications. *Phys. Rev. Lett.* **111**, 024302 (2013).
30. Norris, A. N. Acoustic metafluids. *J. Acoust. Soc. Am.* **125**, 839-849 (2009).
31. Lai, Y. *et al.* Illusion optics: the optical transformation of an object into another object. *Phys. Rev. Lett.* **102**, 253902 (2009).
32. Kan, W. *et al.* Acoustic illusion near boundaries of arbitrary curved geometry. *Sci. Rep.* **3**, 1427 (2013).
33. Silva, A. *et al.* Performing mathematical operations with metamaterials. *Science* **343**, 160-163 (2014).

**Acknowledgements**

This work was partially supported by the Multidisciplinary University Research Initiative grant from the Office of Naval Research (N00014-13-1-0631).

**Author contributions**

Y. X. and S. A. C. conceived the idea. Y. X., C. S. performed the simulation, theoretical analysis and conducted the experiments. Y. X., W. W, J. L. fabricated the sample, B.-I. P. helped with the experiment. Y. X., C. S., S. A. C. and Y. J. prepared the manuscript. All authors contributed to discussions.